\documentclass[fleqn,twoside]{article}
\usepackage{espcrc2}
\usepackage{graphicx}
\usepackage[figuresright]{rotating}
\usepackage{epsfig}

\newcommand{\khzcmq}{\ensuremath{\mathrm {kHz/cm^2}}}

\newcommand{\cmq}{\mathrm{cm^2}}

\newcommand{\NIM}{Nucl. Instr. and Meth. }
\newcommand{\suva}{\ensuremath{\mathrm{C_2H_2F_4}}}
\newcommand{\iso}{\ensuremath{\mathrm{i-C_4H_{10}}}}
\newcommand{\esa}{\ensuremath{\mathrm{SF_6}}}
\newcommand{\Meg}{\ensuremath{\mathrm{M}{\Omega}}}
\newcommand{\VT}{\ensuremath{V_T}}
\newcommand{\VGAP}{\ensuremath{V_{\rm{gap}}}}

\title{New results from an extensive aging test on bakelite Resistive
Plate Chambers}
\author{G. Carboni\address[ROM2]{Universit\`a di Roma ``Tor Vergata''
  and INFN -- Roma II, Via della Ricerca Scientifica 1, I--00133 Roma,
  Italy}, G. Collazuol\address[FIRE]{Universit\`a di Firenze and
  INFN -- Firenze, Via G. Sansone 1, I--50019, Sesto F.no, Firenze,
  Italy}, S. De
  Capua\addressmark[ROM2], D. Domenici\addressmark[ROM2],
  G. Ganis\addressmark[ROM2], R. Messi\addressmark[ROM2],
G. Passaleva\addressmark[FIRE]\thanks{corresponding author},
E. Santovetti\addressmark[ROM2],
M. Veltri\address[URBI]{INFN--Firenze and Universit\`a di Urbino,
V. S. Chiara 27, I--61029 Urbino, Italy}
}

\begin{document}

\begin{abstract}
We present recent results of an extensive aging test, performed at the
CERN Gamma Irradiation Facility on two single--gap RPC prototypes,
developed for the LHCb Muon System. 
With a method based on a model describing the behaviour of an
RPC under high particle flux conditions, we have
periodically measured the electrode resistance $R$
of the two RPC prototypes over three years: we observe a large spontaneous
increase of $R$ with time, from the initial
value of about 2 M$\Omega$ to
more than 250 M$\Omega$. A corresponding degradation of the RPC rate
capabilities, from more than 3 \khzcmq\ to less than 0.15 \khzcmq\ is
also found.
\vspace{1pc}
\end{abstract}

\maketitle

\section{INTRODUCTION}
Resistive Plate Chamber (RPC) detectors have been proposed 
to cover a large fraction (about
48\%) of the Muon System~\cite{TDR,TDR_RPC} of the LHCb
experiment~\cite{TP}. 
The LHCb apparatus covers the forward part of the solid angle and
is therefore subject to a very large particle flux. 
In particular, the particle rates expected in the
Muon System are significantly larger than those expected in the
corresponding sub--detectors in 
ATLAS~\cite{atlas_tdr} and CMS experiments~\cite{cms_tdr}.
In the regions
covered by RPCs, the maximum particle rate is expected to vary
between 0.25 and 0.75 kHz/cm$^2$,
depending mainly on the polar angle. 
To cope with such high rates, LHCb RPCs are operated
in avalanche mode  
\cite{bib:avalanche1,bib:avalanche2,bib:avalanche3}, 
allowing
to obtain a rate capability up to some 
kHz/$\cmq$ \cite{bib:rpc92,bib:rpc99-1,bib:rpc99-2,bib:rpc99-3}. 
In these conditions,
the rate capability of an RPC is determined by the volume resistivity $\rho$
of the electrodes and scales roughly as $1/\rho$. For this reason, RPCs
used in experiments working at high rates are generally built with 
bakelite electrodes, that can be produced with
resistivities as low as $10^9 \; \Omega$cm.

Variations of the electrode resistivity affect directly 
the rate capability of the RPCs:
it is therefore very important to be able to
monitor this parameter during the chamber operation. These variations can be
due to changes in environmental parameters like temperature and humidity
\cite{bib:umid1,bib:umid2}
or to possible aging effects due to operating conditions.

To study these effects, we have devised, in the
framework of the aging studies for the 
LHCb muon chambers, an extensive series of tests
which started in January 2001 
and is expected to last until December 2002,
exploiting the large CERN Gamma Irradiation Facility~\cite{GIF}.

In this paper we present the latest results from these studies. 
Results from a systematic series of measurements of the
resistivity of the bakelite electrodes of two RPC prototypes 
are presented. A study of the RPC rate capabilities is also discussed.

\section{SETUP OF THE AGING TEST}

The aging tests
have been performed at the Gamma Irradiation 
Facility (GIF) at CERN. The GIF is a test area 
where particle detectors are exposed to an adjustable photon flux 
from an intense $^{137}$Cs radioactive source with an activity of 
about 655 GBq. 
A muon beam from the 
SPS accelerator traverses the test area, so that various measurements
can be performed on detectors in presence of the high 
background flux of photons from the source. In our case the 
source was also used to perform an accelerated 
aging test of the detectors \cite{bib:passaleva}. 
A system of remotely
controlled filters allows us to vary the photon rate by four orders of
magnitude. The photon rate has been found to depend from the filter
absorption factor $Abs$ as $\Phi = 1/Abs^{0.7}$
\begin{figure}[!hb]
\begin{center}
\epsfig{file=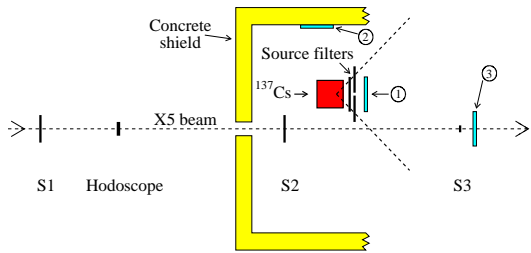,width=0.35\textwidth}
\caption{Schematic view  of the test setup (not to scale). 
The positions of the RPCs 
corresponding to the various measurements are indicated (1-3).
The distance of the RPC from the source in position 1 was about 55 cm,
and in position 3 about 140 cm.  
The scintillator counters (S1-S3 and the Hodoscope) were used 
for measuring the RPC efficiency with the particle beam.}  
\label{fig:setup}
\end{center}
\end{figure}

In our tests, two equal RPC detectors  were 
operated in the GIF area. The detectors ($50 \times 50\; \cmq$
with  a 2 mm gas gap) were built using bakelite plates
(2 mm thick) of nominal
resistivity $\approx 10^{10}\;\Omega$cm, and treated 
internally with linseed oil.

Both detectors were operated with the same gas mixture, 
normally 95\%
$\suva$, 4\% $\iso$ and 1\% $\esa$. 
High-voltage $V_0$, gas mixture composition, currents and temperature were 
continuously monitored and recorded during all the test
period.

The test setup is schematically shown in Figure \ref{fig:setup}.
Normally during the aging test
one detector (RPC A) was placed in position 1, 
very close to the source 
and almost continuously exposed to radiation, 
whereas the second (RPC B) was placed 
far from the source (position 2), to serve as a reference. 

Position 3 was used to perform efficiency measurements with
the particle beam. In this case the signals from the RPCs 
were read out on 3 cm
wide strips using fast electronics. A telescope
of scintillator counters, also shown in Figure \ref{fig:setup},
provided the trigger and a 
hodoscope measured the particle position ($x-y$) with an accuracy
better than 1 cm. At the minimum source attenuation the 
measured flux density in Position 3 
was about 1 kHz/$\cmq$ \cite{bib:ganis}. 

\section{BEHAVIOUR OF RPCs UNDER HIGH FLUX CONDITIONS}

We have observed that the current drawn by RPC detectors subject to a high 
particle flux, shows a
characteristic behaviour:
\begin{itemize}
\item it depends linearly on the applied
voltage, above a certain threshold;
\item it saturates with
increasing flux values;
\item it depends exponentially from the
temperature at a fixed applied voltage. 
\end{itemize}

We have interpreted these
effects with a model, described in details in \cite{bib:carboni}.
In this model we assume that the onset of the avalanche in the
detector arises only above a threshold voltage \VT. Due to space
charge effects, the avalanche charge depends linearly on the effective voltage
\VGAP\ across the gas gap: $q \propto \VGAP-\VT$.
Due to the voltage drop across the RPC electrodes, we have:
$\VGAP=V_0-IR$, where $V_0$ is the nominal applied voltage, $I$ is the
current drawn by the RPC and $R$ is the total volume resistance of the
two electrodes. The current $I$ depends linearly on $q$ through the
incident flux $\Phi$: $I = \Phi q(\VGAP)$. In the limit of infinite flux, 
in order to keep the current finite, we should have $\VGAP = \VT$ and
in these conditions we get a saturation value for the current 
$I_{\rm max} = (V_0-\VT)/R$. At finite flux values, one obtains \cite{bib:carboni}:
\begin{equation}
I = \left(\frac{X}{1+X}\right)\frac{V_0-\VT}{R} =
\frac{V_0-\VT}{R_{\rm{eff}}}\quad , 
\label{eq:model}
\end{equation}
where $X \propto \Phi R$ and $R_{\rm{eff}} = R(1+X)/X$. Equation
\ref{eq:model} is the central equation of the model: it shows that the
current depends indeed linearly on
$V_0$ and that it saturates at high flux values and incorporates the
temperature dependence of the current through the electrode resistance $R$.

\begin{figure}[htb]
\begin{center}
\epsfig{file=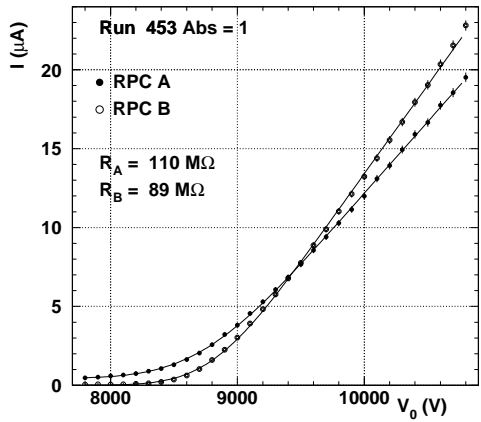,width=0.45\textwidth}
\caption{Current vs. $V_0$  
for RPC A (solid circles) and B 
(open circles) for absorption factor 1. 
The effective resistances $R_{\rm{eff}}$ as obtained from the slopes 
are also shown. In this case we had $X=49.8\pm 8.9$.} 
\label{fig:rfit}
\end{center}
\end{figure}
Typical $I-V_0$ curves of the two 
detectors under test, taken for $Abs=1$
are shown in Figure \ref{fig:rfit}.
A perfect linearity is observed above the threshold voltage \VT\ (about
8500 V in this case). 
The resistance $R$ can thus be measured by fitting these curves
with Equation 
\ref{eq:model} and obtaining the parameter $X$ by measuring the RPC
current at fixed $V_0$ for different flux values. In this way,
the electrode resistance can be measured in a non destructive way and
easily monitored during detector operation.

\section{RESULTS ON RPC RESISTANCE MEASUREMENTS}

In the framework of the RPC aging tests, we have performed a
systematic set of measurements of the RPC electrode resistances using
the method described above.

\begin{figure}[!t]
\begin{center}
\epsfig{file=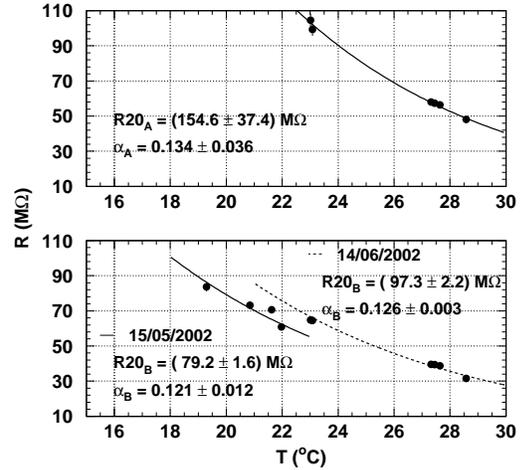,width=0.45\textwidth}
\caption{Resistances of RPC A and B  
plotted versus temperature. The temperature
coefficient $\alpha$ is fitted using 
an exponential dependence \cite{bib:carboni}.}
\label{fig:R_vs_T}
\end{center}
\end{figure}
In order to compare measurements taken at different ambient
temperatures $T$, we have rescaled the values of $R$ to 20$^\circ$C
assuming an exponential dependence from $T$ \cite{bib:carboni}.
Figure \ref{fig:R_vs_T} shows the measurement of $R$ for the two RPCs
at different
temperatures. 
An exponential fit gives an average temperature
coefficient of $\langle\alpha\rangle = 0.126\pm 0.008\; ^\circ {\rm C}^{-1}$
in nice agreement with that obtained for the bakelite slabs
themselves. This confirms that $R$ is indeed the
volume resistance of
the RPC electrodes.
\begin{figure}[!t]
\begin{center}
\epsfig{file=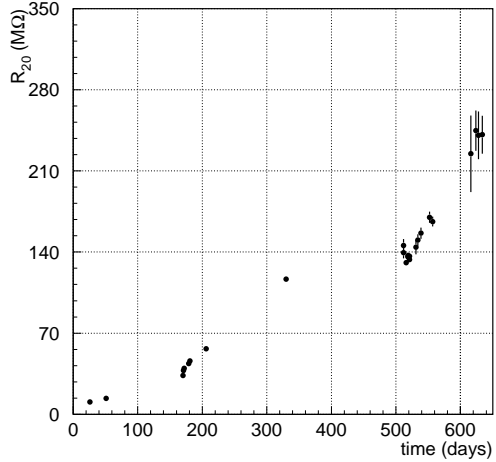,width=0.45\textwidth}
\caption{Measurements of $R$ for RPC A versus time. The time scale
starts at the beginning of the aging test in January
2001} 
\label{fig:res_a}
\end{center}
\end{figure}
\begin{figure}[!t]
\begin{center}
\epsfig{file=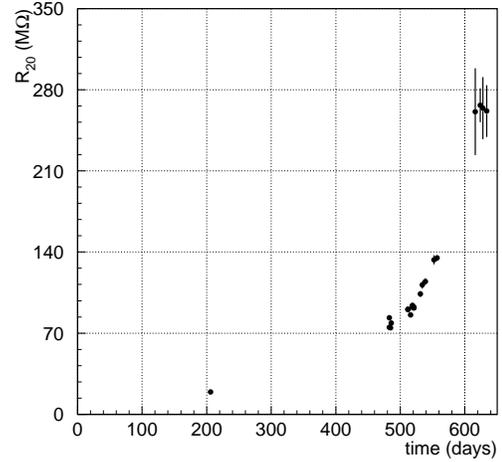,width=0.45\textwidth}
\caption{Measurements of $R$ for RPC B versus time. The time scale
starts at the beginning of the aging test in January
2001} 
\label{fig:res_b}
\end{center}
\end{figure}

\begin{figure}[!b]
\begin{center}
\epsfig{file=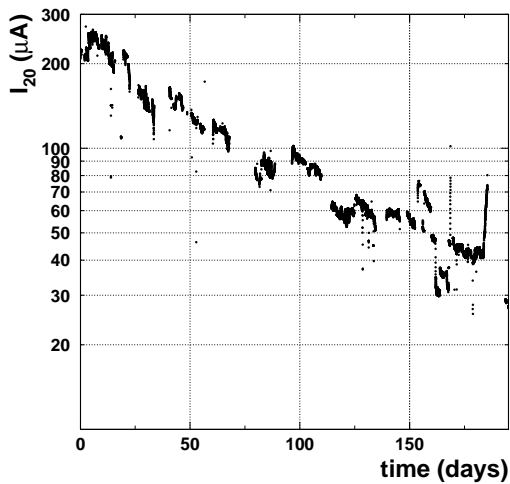,width=0.45\textwidth}
\caption{Current for RPC A corrected for temperature 
(see text) plotted versus time. The detector was placed at
about 55 cm from the source.}
\label{fig:curr}
\end{center}
\end{figure}

The measurements of $R$ rescaled to 20$^\circ$C are shown in
Figure~\ref{fig:res_a} and~\ref{fig:res_b} for RPC A and RPC B respectively.
The time scale starts at the beginning of the aging test in January
2001. In the first 200 days, RPC A was subject to an intense photon
flux and integrated a charge of about 0.4 C/cm$^2$ which corresponds
to about 10 years of operation in LHCb \cite{bib:passaleva}. As shown
in Figure \ref{fig:curr} 
the current drawn by the detector decreased by a factor of six in the
same period. This behaviour was interpreted as a corresponding increase
of the electrode resistance due to the current flowing through the
electrodes themselves. This is indeed clearly visible in
Figure~\ref{fig:res_a}. 
In the
remaining time, the chamber was again operated at the GIF but, because
of the large resistance, the
current drawn by the detector was
negligible.
Still, a large increase of $R$ by
another factor four is observed. This effect is confirmed by the
measurements performed on the prototype B (Figure~\ref{fig:res_b}) 
that was exposed to the
intense photon flux only for a short time. These results seem to
indicate that the
resistivity of the bakelite electrodes tends to spontaneously increase
at a rate much larger than that observed when the RPCs are subject to
the particle flux expected in the LHCb experiment. 
It is clear that this would be by far the main
aging effect over 10 years of operation in LHCb. 

\begin{table*}[!t]
\newcommand{\m}{\hphantom{$-$}}
\newcommand{\pho}{\phantom{0}}
\newcommand{\cc}[1]{\multicolumn{1}{c}{#1}}
\renewcommand{\tabcolsep}{0.8pc} 
\renewcommand{\arraystretch}{1.2} 
\caption{Resistance $R$ and rate capability $r_{max}$ (see text)
measured for RPC
A in three different beam tests at the GIF. The values at 20 $^o$C are
obtained with a temperature coefficient $\alpha = 0.126$ and scaling
$r_{max}$ like $1/R$.}
\begin{tabular}{@{}cccccc}
\hline
Test  & T ($^\circ$C) & $R$ (\Meg) & $R (20^o$C) (\Meg) & $r_{max}$ (\khzcmq) &  $r_{max} (20^o$C) (\khzcmq) \\
\hline
Oct. 1999& 23.0&\pho\pho 1.8 &\pho\pho 2.6 & $> 3$ & $>3$ \\
Aug. 2001& 25.1& \pho 31.6 & \pho 58.3 &   1.1 & 0.6 \\
Jul. 2002& 24.5& 102.4 & 175.7 &   0.4 & 0.2 \\
\hline
\end{tabular}\\[2pt]
\label{tab:ratecap}
\end{table*}

We don't have yet a
quantitative interpretation of these phenomena, although we believe that this
is probably related to a decrease of water content in the bakelite
plates. While water evaporation from the plates is always present, it
is probably enhanced both by the current flowing in the electrodes and
by the flux of dry gas in the chambers. To cross--check
this interpretation we are going to start a series of measurements,
flushing our RPCs with a gas mixture containing water vapour.

\begin{figure}[!t]
\begin{center}
\epsfig{file=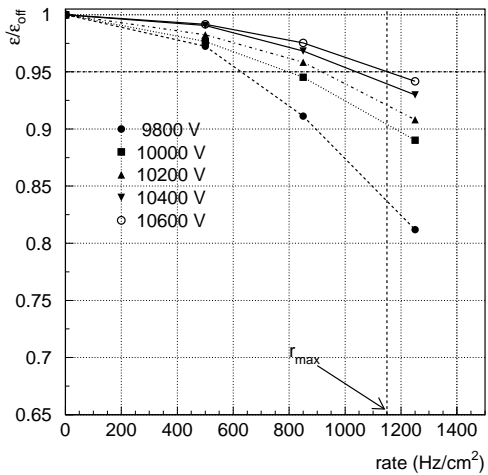,width=0.45\textwidth}
\caption{Efficiency vs. rate for RPC A, at different $V_0$
values, measured in August 2001. The rate capability $r_{max}$ is
defined in the text.
} 
\label{fig:eff_a}
\end{center}
\end{figure}

\section{ANALYSIS OF RATE CAPABILITIES}

The maximum incident particle rate that an RPC can stand is roughly
proportional to the inverse of the electrode volume resistivity. It is
therefore essential to check what is the effect of the increase of the
bakelite resistivity described above on the detector rate
capability. To study quantitatively this effect, we define the rate
capability $r_{max}$ of our prototypes as the maximum rate where their
efficiency is at least 95\%, at a maximum operating voltage of 10.6
kV (this guarantees a plateau of about 400 V below the threshold of
streamer regime), as shown in Figure~\ref{fig:eff_a}. 
The results of three beam tests performed over a
three years period are summarised in Table~\ref{tab:ratecap}. In the
1999 test, the maximum available particle rate at the GIF was about 3
\khzcmq\ but, at this rate, the chamber efficiency was well above
95\%; the rate capability was therefore much larger than 3
\khzcmq. We can
clearly see 
that the rate capability is dropping roughly as $1/R$ as
expected. The rate capability extrapolated to the latest measurements
of $R$ is therefore about 0.15 \khzcmq\ which is well below the rates
foreseen in
LHCb ($> 0.25\; \khzcmq$).
These results have recently brought the LHCb collaboration to abandon
the RPC technology for the Muon System of the experiment.

\section{CONCLUSIONS}

We have developed and applied a method to measure RPC electrode
resistance $R$, on--line, during  chamber operations, in an easy and non 
destructive way. Using this method we have studied extensively the
aging effects on bakelite RPCs. We have observed an increase of $R$
by two orders of magnitude in 
about three years. Pure spontaneous aging is the dominant effect.
 The RPC rate capability dropped correspondingly
from a few \khzcmq\
to less than 0.15 \khzcmq. These results brought the LHCb
collaboration to abandon the RPC technology for the Muon System of the
experiment.

\end{document}